\documentstyle[12pt]{article}
\begin{document}
\begin{titlepage}
\hfill\begin{tabular}{l}HEPHY-PUB 692/98\\UWThPh-1998-39\\July 1998
\end{tabular}\\[2cm]
\begin{center}
{\Large\bf SPINLESS SALPETER EQUATION:}\\[2ex]
{\Large\bf ANALYTIC RESULTS}\\[2.5cm]
{\Large\bf Wolfgang LUCHA}\\[.5cm]
Institut f\"ur Hochenergiephysik,\\
\"Osterreichische Akademie der Wissenschaften,\\
Nikolsdorfergasse 18, A-1050 Wien, Austria\\[1cm]
{\Large\bf Franz F. SCH\"OBERL}\\[.5cm]
Institut f\"ur Theoretische Physik,\\
Universit\"at Wien,\\
Boltzmanngasse 5, A-1090 Wien, Austria\\[1.7cm]
\end{center}
\begin{abstract}
The spinless Salpeter equation is the combination of relativistic kinematics
with some static interaction potential. The nonlocal nature of the
Hamiltonian resulting from this approximation renders difficult to obtain
rigorous analytic statements on resulting solutions. In view of this
unsatisfactory state of affairs, we calculate analytic upper bounds on the
involved energy levels, and, for the Coulomb potential, the ground-state
energy at the critical coupling constant.
\end{abstract}
\vspace*{1cm}

\normalsize\it
Invited talk, presented by F. F. Sch\"oberl, at the ``XI International
Conference: Problems of Quantum Field Theory,'' at the Joint Institute for
Nuclear Research, Bogoliubov Laboratory of Theoretical Physics, July 13--17,
1998, Dubna, Russia.
\end{titlepage}

\section{Introduction: The Spinless Salpeter Equation}

The most straightforward generalization of standard nonrelativistic quantum
theory towards reconciliation with requirements imposed by special relativity
is to describe a quantum system under consideration by the well-known {\em
spinless Salpeter equation}. Consider a quantum system the dynamics of which
is governed by a---by assumption self-adjoint---Hamiltonian $H$ of the form
\begin{equation}
H=T+V\ ,
\label{eq:ham-sseq}
\end{equation}
where $T$ denotes the square-root operator of the relativistic expression for
the kinetic energy of some particle of mass $m$ and momentum ${\bf p}$, that
is, $T\equiv\sqrt{{\bf p}^2+m^2}$,~and $V=V({\bf x})$ represents an arbitrary
coordinate-dependent static interaction potential. The eigenvalue equation
for this Hamiltonian $H$, $H|\chi_k\rangle=E_k|\chi_k\rangle$,
$k=0,1,2,\dots$,~for Hilbert-space eigenvectors $|\chi_k\rangle$
corresponding to energy eigenvalues
$$
E_k\equiv\frac{\langle\chi_k|H|\chi_k\rangle}{\langle\chi_k|\chi_k\rangle}\ ,
$$
is the one-particle spinless Salpeter equation. However, because of the
nonlocality~of $H$, it is extremely hard to obtain analytic statements from
this equation of motion.

Of particular importance in physics are ``central'' potentials, i.e.,
potentials which depend only on the radial coordinate $r\equiv|{\bf x}|$.
Among these, the most prominent one is the Coulomb potential $V_{\rm C}(r)$,
which is parametrized by some coupling constant~$\alpha$:
\begin{equation}
V({\bf x})=V_{\rm C}(r)=-\frac{\alpha}{r}\ ,\quad\alpha>0\ .
\label{eq:coulpot}
\end{equation}
The semirelativistic Hamiltonian (\ref{eq:ham-sseq}) with the Coulomb
interaction potential $V$ in~(\ref{eq:coulpot}) defines the {\em spinless
relativistic Coulomb problem}. In the past, the interest in this has
undergone an eventful history. (For a rather comprehensive review, consult
Ref.~\cite{lucha94}.)

By examining \cite{herbst77} the spectral properties of the Hamiltonian
(\ref{eq:ham-sseq}) with the potential (\ref{eq:coulpot}) one infers the
existence of a Friedrichs extension up to the critical value $\alpha_{\rm
c}=2/\pi$ of $\alpha$. As far as analytic statements about this spinless
relativistic Coulomb problem, in particular, about the resulting energy
levels, are concerned, up to now one has to be content with few series
expansions of the energy eigenvalues in powers of $\alpha$
\cite{leyaouanc94,brambilla95}, which then are, of course, only significant
for some region of rather small values of~$\alpha$.

\section{Variational Upper Bounds on Energy Levels}

Variational upper bounds on the eigenvalues of the above Hamiltonian $H$ are
derived rather easily in the following way \cite{lucha96upper}. We introduce
an arbitrary real parameter $\mu$~and make use of the positivity of the
square of the (obviously self-adjoint) operator $T-\mu$,
$$
0\le (T-\mu)^2=T^2+\mu^2-2\,\mu\,T={\bf p}^2+m^2+\mu^2-2\,\mu\,T\ ,
$$
in order to find, for the kinetic energy $T$, the operator inequality (see
also Ref.~\cite{martin88})
$$
T\le\frac{{\bf p}^2+m^2+\mu^2}{2\,\mu}\quad\mbox{for all}\ \mu>0\ ,
$$
and, therefore, for the semirelativistic Hamiltonian $H$ in
(\ref{eq:ham-sseq}), the operator inequality
$$
H\le\widehat H_{\rm S}(\mu)\quad\mbox{for all}\ \mu>0\ ,
$$
with some Schr\"odinger-like Hamiltonian, denoted by $\widehat H_{\rm
S}(\mu)$ and given by the unusual form
$$
\widehat H_{\rm S}(\mu)=\frac{{\bf p}^2+m^2+\mu^2}{2\,\mu}+V\ .
$$
Now, invoking the min--max principle, we infer that the set of energy
eigenvalues~$E_k$, $k = 0,1,2,\dots$, of the Hamiltonian $H$, if ordered
according to $E_0 \le E_1 \le E_2 \le \dots$,~is bounded from above by the
corresponding set of energy eigenvalues $\widehat E_{{\rm S},k}(\mu)$ of
$\widehat H_{\rm S}(\mu)$, if the latter are similarly ordered according to
$\widehat E_{{\rm S},0}(\mu)\le\widehat E_{{\rm S},1}(\mu)\le\widehat E_{{\rm
S},2}(\mu)\le\dots$,~i.e., $E_k\le\widehat E_{{\rm S},k}(\mu)$ for all
$\mu>0$, and is, consequently, also bounded by the minimum of~all these
energy eigenvalues:
$$
E_k\le\min_{\mu>0}\widehat E_{{\rm S},k}(\mu)\ .
$$

For the Coulomb potential (\ref{eq:coulpot}), these energy eigenvalues
$\widehat E_{{\rm S},n}(\mu)$ are given by~\cite{lucha96upper}
$$
\widehat E_{{\rm S},n}(\mu)=\frac{1}{2\,\mu}
\left[m^2+\mu^2\left(1-\frac{\alpha^2}{n^2}\right)\right]\ ,
$$
where the total quantum number $n$ is defined according to $n=n_{\rm
r}+\ell+1$. Minimizing the latter expression with respect to the parameter
$\mu$, we obtain as minimal bound~\cite{lucha96upper}
\begin{equation}
\min_{\mu>0}\widehat E_{{\rm S},n}(\mu)=m\,\sqrt{1-\frac{\alpha^2}{n^2}}\ .
\label{eq:varbound}
\end{equation}

Comparing with the results obtained in the framework of perturbation
theory~\cite{brambilla95}, which read, for instance, for the level $n=2$,
$\ell=1$,
$$
\frac{E}{m}=1-\frac{\alpha^2}{8}-\frac{7\,\alpha^4}{384}+
\frac{727\,\alpha^6}{82944}+\dots\ ,
$$
we observe that our upper bound (\ref{eq:varbound}) not only possesses the
elegance of a particularly simple form but is also valid for all
$\alpha\le\alpha_{\rm c}$ and arbitrary levels of excitation. A~quick glance
at Table~\ref{tab:compare} shows that the relative error of our bound is
always less than 0.1~$\%$.

\begin{table}[htb]
\caption[$\quad$]{Comparison of the perturbatively computed
\cite{brambilla95} energy eigenvalues for, e.g., the level $n=2$ and $\ell=1$
with~our variational upper bounds, extracted from
Eq.~(\ref{eq:varbound})}\label{tab:compare}
\vspace*{.5ex}
\begin{center}
\begin{tabular}{lll}
\hline\hline
&&\\[-1.5ex]
\multicolumn{1}{c}{$\alpha$}&\multicolumn{2}{c}{$E/m$}\\[1ex]
\multicolumn{1}{c}{}&\multicolumn{1}{c}{\hspace{1ex}Perturbation
Theory \cite{brambilla95}\hspace{1ex}}&\multicolumn{1}{c}
{\hspace{1ex}Variational Procedure, Eq.~(\ref{eq:varbound})\hspace{1ex}}\\[1ex]
\hline
&&\\[-1.5ex]
$\quad$0.0155522$\quad$&$\quad$0.999969765$\quad$&$\quad$0.999969766$\quad$\\
$\quad$0.1425460$\quad$&$\quad$0.997452$\quad$&$\quad$0.997457$\quad$\\
$\quad$0.2599358$\quad$&$\quad$0.99147$\quad$&$\quad$0.99152$\quad$\\
$\quad$0.3566678$\quad$&$\quad$0.9838$\quad$&$\quad$0.9840$\quad$\\
$\quad$0.4359255$\quad$&$\quad$0.975$\quad$&$\quad$0.976$\quad$\\
$\quad$0.5$\quad$&$\quad$0.967$\quad$&$\quad$0.9682$\quad$\\[1ex]
\hline\hline
\end{tabular}
\end{center}
\end{table}

\section{The Energy at the Critical Coupling Constant}

Let us apply again the min--max principle. Our particular (somewhat
sophisticated) choice for the basis vectors $|\psi_k\rangle$ to be adopted
here is defined by trial functions $\psi_k(r)$ given in configuration-space
representation (with the radial coordinate $r\equiv|{\bf x}|$)
by~\cite{lucha96num}
$$
\psi_k(r)=\sqrt{\frac{(2\,m)^{2\,k+2\,\beta+1}}{4\pi\,\Gamma(2\,k+2\,\beta+1)}}
\,r^{k+\beta-1}\exp(-m\,r)\ ,\quad\beta\ge 0\ ,\quad m>0\ .
$$
Whereas $k$ indicates a positive integer, $k = 0,1,2,\dots$, the parameter
$\beta$ is introduced to allow, for some given value of the coupling constant
$\alpha$, of a total cancellation of the divergent contributions to the
expectation values of kinetic energy $T$ and interaction potential $V_{\rm
C}(r)$: $\beta=\beta(\alpha)$. More precisely: in order to provide for that
cancellation, the parameter $\beta$ has to be adjusted for the ground state
according to the relation~\cite{raynal94}
$$
\alpha=\beta\cot\left(\frac{\pi}{2}\,\beta\right)\ ,
$$
which implicitly determines $\beta$ as a function of the coupling constant
$\alpha$. In particular, this relation tells us that the critical coupling
constant, $\alpha_{\rm c}$, is approached for $\beta\to 0$. From the
behaviour of the matrix elements $\langle\psi_i|T|\psi_j\rangle$ of the
kinetic energy $T$ for~large momenta $p$ and of the matrix elements
$\langle\psi_i|V_{\rm C}(r)|\psi_j\rangle$ of the Coulomb potential $V_{\rm
C}(r)$ at small distances $r$, respectively, it should become evident that,
for our particular choice of basis states $|\psi_k\rangle$, these
singularities can arise only in matrix elements taken with respect to the
ground state $|\psi_0\rangle$, that is, merely in
$\langle\psi_0|T|\psi_0\rangle$ and $\langle\psi_0|V_{\rm
C}(r)|\psi_0\rangle$.

The remainder of our way is straightforward. With the above basis
functions,~we obtain, by calculating the $2\times 2$ energy matrix (and
performing very carefully the~limit $\beta\to 0$), for the roots of the
corresponding characteristic equation the expression~\cite{critical96}
$$
\frac{\widehat E}{m}=\frac{2}{15\,\pi}
\left(60\ln 2-23\,\pm\sqrt{(60\ln 2)^2-4800\ln 2+1649}\right)\ ,
$$
which yields the upper bound $\widehat E_0/m=0.484288\dots$ for the
ground-state eigenvalue of the Coulombic Hamiltonian. In principle, the $d$
roots of the characteristic equation may be determined algebraically up to
and including the case $d=4$. For $d=4$,~our method constrains the
lowest-lying energy level by the bound $\widehat E_0/m=0.4842564\dots$ On the
other hand, at the critical coupling constant $\alpha_{\rm c}$ the best
numerically obtained bounds are given by \cite{raynal94}
$$
0.4825\le\frac{E_0}{m}\le 0.4842910\quad\mbox{for}\ \alpha=\alpha_{\rm c}\ .
$$
Comparing these only numerically determined bounds on the ground-state
energy~$E_0$ with the above upper bound, we realize that even our $2\times 2$
analytical bound lies~well within the numerically obtained range.
Consequently, we obtain a clear improvement of the best upper bound available
up to now for the ground-state energy level of the spinless relativistic
Coulomb problem at the critical value of the coupling constant.


\begin{thebibliography}{9}
\bibitem{lucha94} W.~Lucha and F.~F.~Sch\"oberl, {\it All Around the Spinless
Salpeter Equation}, in: {\it Proceedings of the International Conference on
Quark Confinement and the Hadron Spectrum} (Como, Italy, June 1994), eds.
N.~Brambilla and G.~M.~Prosperi (World Scientific, River Edge, N.~J., 1995)
p.~100.
\bibitem{herbst77} I.~W.~Herbst, Commun.~Math.~Phys. {\bf 53} (1977) 285;
{\bf 55} (1977) 316 (addendum).
\bibitem{leyaouanc94} A.~Le Yaouanc et al., Ann.~Phys.~(N.~Y.) {\bf 239}
(1995) 243.
\bibitem{brambilla95} N.~Brambilla and A.~Vairo, Phys.~Lett.~B {\bf 359}
(1995) 133.
\bibitem{lucha96upper} W.~Lucha and F.~F.~Sch\"oberl, Phys.~Rev.~A {\bf 54}
(1996) 3790; hep-ph/9603429.
\bibitem{martin88} A.~Martin, Phys.~Lett.~B {\bf 214} (1988) 561.
\bibitem{lucha96num} W.~Lucha and F.~F.~Sch\"oberl, Phys.~Rev.~A {\bf 56}
(1997) 139; hep-ph/9609322.
\bibitem{raynal94} J.~C.~Raynal et al., Phys.~Lett.~B {\bf 320} (1994) 105.
\bibitem{critical96} W.~Lucha and F.~F.~Sch\"oberl, Phys.~Lett.~B {\bf 387}
(1996) 573; hep-ph/9607249.
\end{thebibliography}
\end{document}